\documentclass{elsart1p}

\usepackage{bbm}
\usepackage{amssymb}
\usepackage{amsmath}

\newcommand{\nc}{\newcommand}
\nc{\be}{\begin{equation}}
\nc{\ee}{\end{equation}}
\nc{\bea}{\begin{align}}
\nc{\eea}{\end{align}}
\nc{\nn}{\nonumber}
     
\nc{\lp}{\left(}
\nc{\rp}{\right)}

\catcode`\*=11  
\def\slashsym#1#2{\mathpalette{\sl*sh{#1}}{#2}}
\def\sl*sh#1#2#3{\ooalign{\setbox0=\hbox{$#2\not$}
                          $\hfil#2\mkern-24mu\mkern#1mu
                           \raise.15\ht0\box0\hfil$\cr
                          $#2#3$}}
\catcode`\*=12

\def\pslash{{\slashsym8p}}

\def\Aslash{{\slashsym9A}}

\def\Bslash{{\slashsym6B}}

\def\Ac{{\mathcal{A}}}

\def\Fc{{\mathcal{F}}}
\def\Hc{{\mathcal{H}}}

\def\Dc{{\mathcal{D}}}

\def\Oc{{\mathcal{O}}}

\def\Det{{\mathrm{Det\,}}}

\nc{\markx}{$ \clubsuit $}

\nc{\eq}{{Eq.}}
\nc{\eqs}{{Eqs.}}
\nc{\tr}{\textrm{tr}}

\begin{document}
\begin{frontmatter}


\title{
Are CP Violating Effects in the Standard Model Really Tiny?
}

\author[Andres]{A.~Hernandez\corauthref{eAH}},
\author[Konstandin]{T.~Konstandin\corauthref{eTK}},
\author[Andres]{M.~G.~Schmidt\corauthref{eMS}}
\address[Andres]{Institut f\"ur Theoretische Physik, Heidelberg
University, Germany}
\address[Konstandin]{Institut de F\'isica d'Altes Energies,
Universitat Aut\`onoma de Barcelona, Spain}
\corauth[eAH]{A.Hernandez@thphys.uni-heidelberg.de}
\corauth[eTK]{Konstand@ifae.es}
\corauth[eMS]{\hspace{0.25cm}M.G.Schmidt@thphys.uni-heidelberg.de}

\date{\today}

\begin{abstract}
We derive an effective action of the
bosonic sector of the Standard Model by integrating out the fermionic
degrees of freedom in the worldline approach. The CP violation due to the
complex phase in the CKM matrix gives rise to CP-violating operators in the
effective action.  We calculate the prefactor of the appropriate next-to-leading order
operators and give general estimates of CP violation in the bosonic
sector of the Standard Model. In particular, we show that the
effective CP violation for weak gauge fields is not suppressed by the
Yukawa couplings of the light quarks and is much larger than the bound
given by the Jarlskog determinant.
\end{abstract}

\maketitle

%
%
\end{frontmatter}
\section{Motivation and Introduction}
CP-violating effects in the Standard Model stem from the Yukawa couplings of the quarks, and
are generally very small. They are due to the special flavor structure of the Standard Model~\cite{CKMmatrix:1973,Jarlskog:1985cw}.
To be explicit, the CP violation arises from the following terms in the Lagrangian
\be
 Y^u_{ij}  \, \bar Q_L^i \, u_R^j \phi + 
Y^d_{ij}  \, \bar Q_L^i \, d_R^j \tilde\phi +{h.c.},
\ee
where $Q_L$ denotes the left-handed quark $SU(2)_L$ doublet, $d_R$ and
$u_R$ denote the right-handed quark singlets and $\phi$ denotes the
Higgs doublet.  We also defined the field $\tilde \phi$ by
\be
\tilde \phi = \epsilon \phi^* = 
\begin{pmatrix}
0 & -1 \\
1 & 0 \\
\end{pmatrix}
\begin{pmatrix}
\phi^0 \\
\phi^+ \\
\end{pmatrix}^*=
\begin{pmatrix}
-\phi^- \\
\phi^{0*} \\
\end{pmatrix},
\ee
and $Y^u$ and $Y^d$ denote the Yukawa coupling matrices. Under CP
conjugation, the Yukawa couplings transform as
\be
{\cal CP} \, Y^{u/d} \, {\cal CP}^{-1} = \left(Y^{u/d}\right)^{*},
\ee
such that imaginary entries in $Y^{u/d}$ potentially constitute CP
violation. Spontaneous breakdown of the $SU(2)_L$ symmetry gives then
rise to the SM quark masses. However, not all entries in the Yukawa matrices are observable.  The
Yukawa couplings are the only terms in the SM Lagrangian that are
sensitive to global $SU(3)_R$ flavor transformations. This leads to the
conclusion that physical observables can only depend on the combinations
$m_u m_u^\dagger$ and $m_d m_d^\dagger$. In addition, there are six
global phases in the left-handed quark sector that are unobservable in
the SM.

In Ref.~\cite{Jarlskog:1985prl} it was shown that in perturbation
theory the first CP odd combination of the Yukawa couplings that is
invariant under these transformations is the so-called Jarlskog
determinant
\be
\label{eq_def_Jinv}
\delta_{CP} = \textrm{Im  Det} \left[ m_u
m_u^\dagger, m_d
m_d^\dagger\right] 
= J \prod_{i<j} \frac{\tilde m_{u,i}^2 - \tilde m_{u,j}^2}{v^2}
\prod_{i<j} \frac{\tilde m_{d,i}^2 - \tilde m_{d,j}^2}{v^2}\simeq 10^{-19},
\ee
where $J=s_1^2s_2s_3c_1c_2c_3 \sin(\delta)=(3.0\pm0.3)\times10^{-5}$,
%
%
and $\tilde m^2_{u/d}$ denotes the diagonalized mass matrices
according to
\be
m_d m_d^\dagger = D \tilde m^2_d D^\dagger, \quad
m_u m_u^\dagger = U \tilde m^2_u U^\dagger.
\ee
The Jarlskog determinant in Eq.~(\ref{eq_def_Jinv}) reflects the fact
that CP violation is absent if any two up masses or any two down
masses are equal. This is required since in this case there is an
additional global flavor symmetry that can be used to remove all
complex phases from the Yukawa matrices (in the SM case of three quark
families). However, the above argument is based on the assumption that the
observable under consideration is perturbative in the Yukawa
couplings.

A baryogenesis mechanism that is based on the SM would be
most compelling, but this requires that the Jarlskog determinant as an
upper bound on CP violation is evaded. In principle, there are several 
possibilities to avoid this dilemma and to obtain a significant source of 
CP violaton in the SM as required by baryogenesis. For
example, during a first-order phase transition, the Higgs vev changes
and hence makes it possible to construct rephasing invariants that do
not only contain the masses but also their derivatives that are
non-vanishing during the phase
transition~\cite{Konstandin:2003dx}.

The possibility that 
concerns us here is dealing with the complicated non-perturbative dynamics by integrating out 
the fermions and constructing an effective action of the bosonic variables only. 
CP violation then appears in higher-dimensional terms of an effective Lagrangian. Main 
motivation for this approach is the scenario of cold electroweak baryogenesis
\cite{Smit:2003,Smit} that specifically utilizes
lattice simulations of the bosonic sector of the SM with higher
dimensional operators that violate the CP symmetry.

\section{Effective Action}
We are interested then in the following one-loop effective action

\be\label{WE}
-W [ \Phi,\Pi,A,B,K ] = \log \Det [\pslash - i \Phi(x) - \gamma_5 \Pi(x) - \Aslash(x) - \gamma_5 \Bslash(x)].
\ee
The solution is obtained for outer fields with a general internal group structure, e.g. a flavour matrix structure. 
Only later do we specialize to the SM. We analyze the real and imaginary parts of the effective action separately, 
and we are interested in the imaginary part that contains the CP-violating contributions to the action
\be
\label{Sep}
-W^{+}-i\,W^{-}=\log \left(\vert \Det[\Oc]\vert\right)+i\,\arg\left(\Det[\Oc]\right).
\ee

To obtain the effective action we work with a worldline representation
of the chiral current for which a manifestly chiral covariant
expression exists. This current can then be integrated to obtain the
effective action~\cite{worldline, worldline2, Salcedo1}.  This integration
rather proceeds by matching: First, a general effective action is
proposed, which has the expected chiral and covariant properties. The
functional variation of this action is then matched to the covariant
current that is obtained using the worldline formalism.  This method
has the advantage that it is both gauge and chiral invariant at each
stage of the calculation. The anomaly only leads to additional
complications in the matching procedure of the lowest order
contributions.


As shown elegantly in Ref.~\cite{Gagne}, the imaginary part of the
action can be reformulated in terms of variables that have a
well-defined behaviour under chiral transformations, namely
%
\begin{align}
\Ac_{\mu}=&
\begin{pmatrix}
A_{\mu}^L & 0 \cr 0 & A_{\mu}^R \\
\end{pmatrix}=
\begin{pmatrix}
A_{\mu}+B_{\mu} & 0 \cr 0 & A_{\mu}-B_{\mu} \\
\end{pmatrix}\nn\\ 
\Hc =&
\begin{pmatrix}
0 & i\, H \\
-i\, H^{\dagger} & 0 \\
\end{pmatrix}=
\begin{pmatrix}
0 & i\, \Phi + \Pi \\
-i\, \Phi + \Pi & 0 \\
\end{pmatrix}.
\end{align}

Next, the derivative expansion of the heat kernel is used. In the
derivative expansion terms are classified by the number of covariant
indices that they carry, so that $\Dc_{\mu}\Hc$ is of first order,
while $\Fc_{\mu\nu}$ is of second order. The worldline formalism is
well suited for this expansion for the following reasons: first, 
we can avoid $\gamma-$matrix algebra; second, the momentum integration is
omitted and replaced by the rather trivial integration in $\tau$
space; third, the result is expressed in $x-$space; and last the 
method is easily implemented with computer algebra.

\section{NLO Results}
The effective action is most compactly presented in the labeled
operator notation that was introduced in Ref.~\cite{Salcedo1}, and
used also in Ref.~\cite{worldline}. In this notation, mass matrices
obtain an additional subscript that indicates the position of the mass
matrix in a subsequent product of operators. For example, using
this notation we write
\be
m_1 m^3_2 m^2_3 \, \Dc_{\mu}\Hc \, \Dc_{\nu}\Hc =
m \, \Dc_{\mu}\Hc \, m^3 \,  \Dc_{\nu}\Hc \, m^2.
\ee
A detailed definition and applications of this notation can be found
in Ref.~\cite{Salcedo1} and we refer the reader to this work. 

The covariant effective action in 4-dimensions and to leading order in the covariant derivative expansion
looks like

\be
W_c^{-}=\epsilon^{\mu\nu\lambda\sigma}\left\langle N_{123}\Dc_{\mu}\Hc\Dc_{\nu}\Hc\Fc_{\lambda\sigma}
+N_{1234}\Dc_{\mu}\Hc\Dc_{\nu}\Hc\Dc_{\lambda}\Hc\Dc_{\sigma}\Hc \right\rangle.
\ee

It is clear that there can be no CP-violating contribution from this expression\cite{Smit} and we must then
look for contributions at next-to-leading order. 

Using the method developed in Ref.~\cite{worldline} we
calculated the effective action explicitly in next-to-leading order~\cite{worldline2}. The explicit form is not shown here for
space concerns. Interestingly, there is only one contribution to the
CP-violating part of the effective action, namely
\be\label{eq:CPterm}
\frac{1}{8(4\pi)^2}\frac{3}{16} \frac{J\,\kappa^{CP}}{\tilde m_c^2}
\epsilon^{\mu\nu\lambda\sigma}\int d^4x
\biggl(Z_{\mu}W^{+}_{\nu\lambda}W^{-}_{\alpha}
\left(W^{+}_{\sigma}W^{-}_{\alpha}+W^{+}_{\alpha}W^{-}_{\sigma}\right)
+ \, c.c.\biggr),
\ee
with
\be
\kappa^{CP} \approx 9.87.
\ee

Finally, notice that the action can always be rewritten in $SU(2)_L$
gauge invariant quantities. For example, the charged gauge fields can
be rewritten as
\be
W_{\mu\nu}^+ = \frac{\phi^\dagger W_{\mu\nu} \tilde \phi}{ \phi^\dagger \phi}, \quad
W_{\mu\nu}^- = \frac{{\tilde \phi}^\dagger W_{\mu\nu} \phi}{ \phi^\dagger \phi}, \quad
W_{\mu}^+ = \frac{\phi^\dagger {\cal D}_\mu \tilde \phi}{ \phi^\dagger \phi}, \quad
W_{\mu}^- = \frac{{\tilde \phi}^\dagger {\cal D}_{\mu} \phi}{ \phi^\dagger \phi}.
\ee
%
\section{Conclusions\label{sec_concl}}

We calculated the CP-violating contributions to the effective action
in the bosonized Standard Model in next-to-leading order in the
gradient expansion. The resulting action should be
valid at least for bosonic fields whose energy scale does not exceed much the charm
mass. This observation is based on the fact that the action after IR
regularization remains finite in the limit of vanishing up and down
quark masses. We find that the coefficients of the resulting
dimension-six operators are suppressed by the charm mass and the
Jarlskog invariant $J$ but are many orders larger than the Jarlskog
determinant $\delta_{CP}$.
\section*{Acknowledgments}

T.K. is supported by the EU FP6 Marie Curie Research \& Training
Network 'UniverseNet' (MRTN-CT-2006-035863).  A.H. is supported by
CONACYT/DAAD, Contract No.~A/05/12566.

\end{document}